\documentclass[fleqn,12pt,twoside]{article}
\usepackage{espcrc1}

\usepackage{graphicx}
\usepackage[figuresright]{rotating}

\newcommand{\AmS}{{\protect\the\textfont2
  A\kern-.1667em\lower.5ex\hbox{M}\kern-.125emS}}

\title{
Non-identical particle correlation analysis as a probe of transverse flow}

\author{F. Reti\`ere \address[LBL]{Lawrence Berkeley
National Laboratory}
 for the STAR collaboration
        \thanks{For the full author list and acknowledgements, see Appendix
"Collaborations" of this volume}
}
       
\begin{document}

\maketitle

\begin{abstract}

Non-identical two particle correlation functions probe
asymmetries between the average space-time emission points of different particle species.
The system collective expansion would produce such asymmetry because
massive particles, such as protons, are on average more pushed towards
the edge of the system, than lighter ones, i.e. pions. Measuring pion-kaon, pion-proton and kaon-proton correlation functions, using the STAR detector, we show that the data are consistent with the effect of transverse flow in Au-Au collisions at $\sqrt{s_{NN}} = 130$ GeV.

\end{abstract}

\section{Transverse expansion in ultra-relativistic heavy ion collisions}

High energy densities are reached in ultra-relativistic heavy ion collisions~\cite{PhenixEt},
which may lead to a transition from hadronic matter to partonic matter.
In order to assess the formation
of a partonic state of matter, which requires  forming a large volume of interacting
matter, it is mandatory to show evidences that the particles emerging
from the system exhibit collective behavior.
While it is not a required condition, showing that the system reaches thermal equilibrium would provide
such clue. Models that assume chemical equilibrium at hadronization
describe very well the relative abundance of most particle species~\cite{ThMod}. 
However, such 
models are  also successful at describing yields in $p-p$ and $e^{+}-e^{-}$ collisions
where thermalization is not achieved.
Another clue, may be obtained from studying collective expansion of the system
which must arise if particles interact with one another. Indeed, a pressure 
gradient would then build up due to the large initial energy density
gradient found between the core and the edge of the system. 
Conclusive evidence of the system's collective expansion have been 
gathered from several independent analysis: elliptic flow, transverse mass
spectra and two-particle correlations~\cite{CollExp}.  These measurements 
in Au--Au collisions at $\sqrt{s_{NN}} = 130$ GeV are consistently interpreted 
in the blast wave framework relying on transverse flow, i.e. collective
expansion in the transverse plane. The blast wave parameterization
is described in Ref~\cite{BW}. Dealing with central events, cylindrical symmetry is 
assumed, and the calculations
parameters are: temperature ($T=110$ MeV),  transverse flow
rapidity ($\rho(r) =  3/2 \langle \rho \rangle (r/R)$ with $\langle \rho \rangle = 0.6$),  radius of the system ($R = 13$ fm), and 
emission duration ($\Delta\tau = 1.5$ fm/c) . $T$ and $\langle \rho \rangle$ are 
extracted from pion, kaon and proton transverse mass spectra. 
$R$ and $\Delta\tau$ are extracted from the pion source radii, $R_{out}$ and $R_{side}$.

The interpretation of data in terms of transverse flow may not be unique.  For example, the increase of the mean transverse momentum with particle mass could be explained by initial state scattering ~\cite{InitRescat}. Since spatial separation between
sources of different species can be extracted from non-identical particle correlation functions, they provide an independent cross-check of the transverse flow prescription.
Indeed, in the blast wave framework, the competition between thermal motion and transverse flow leads to a significantly different emission pattern depending
on the particle transverse mass. When thermal motion dominates, i.e. for  low transverse momentum pions, the emission points of particles going in a 
specific direction are spread over a large fraction of the source. On the other hand, this correlation is strong for particles with larger transverse mass, and their
emission points are more likely to be shifted along the particle momentum, towards
the edge of the system.
Thus, transverse flow implies that
the average transverse radii where particles are emitted rise with their transverse mass.
Pions are emitted the closest to the center of the system, then kaons, then protons. This ordering may be different in hadronic cascade models where hadronic crosss sections matter.

\section{Non-identical particle correlation functions at STAR}

The correlation 
between non-identical particles arises from Coulomb and nuclear interactions.
It depends  on $\overrightarrow{k}^*$ the relative momentum and 
$ \overrightarrow{\Delta r}^*$ the relative separation in the pair rest frame.
The correlation is strong, when $|\overrightarrow{k}^*| = k^*$ and 
$| \overrightarrow{\Delta r}^*|$ are small\footnote{We note that small $k^*$
implies that particles have the same velocity and not the same momentum.}. 
In Ref. ~\cite{NonIdMeth} it was noticed that 
if the space-time emission points of two different particle species
do not coincide, the correlation strength due to the Coulomb interaction, and in some cases due to the nuclear interaction, 
depends on whether both 
particles move towards each other (stronger correlation) or away from 
each other (weaker correlation).  To differentiate both cases we calculate 
$k^*_{out}$, the projection of $\overrightarrow{k}^*$ along the pair
transverse momentum. By convention, we calculate $k^*_{out}$ for the lighter
particle. Thus, $k^*_{out}>0$ ($k^*_{out}<0$) means that the lighter particle transverse velocity is larger (smaller) than the heavier one's. Thus, if, as predicted in the blast wave framework
pions are emitted
closer to the center of the source than kaons, the pion-kaon correlation will be stronger when $k^*_{out}>0$ than
when $k^*_{out}<0$, since in the first case pions tend to catch up with kaons
while in the second case they tend to move away from kaons. 

Non-identical particle correlations are studied by
constructing correlation functions  : 
$C_{2}(x) = A(x) /  B(x)$ where
$x = k^{*}sign(k^*_{out})$, 
$A$ denotes the distribution for pairs of particles from the same event, and $B$ the distribution for pairs of particle from different events. 
Pions, kaons and protons are reconstructed and identified using the STAR
Time Projection Chamber, in  Au-Au collisions at $\sqrt{s_{NN}} = 130$
 GeV and $\sqrt{s_{NN}} = 200$ GeV.  
Central events accounting for 12\% of the total cross section are selected. 
Clean pion, kaon and proton samples are selected using
their specific energy loss ($dE/dx$).  This selection limits the acceptance of pions 
to $0.08 < p_t < 0.6$ GeV/c, of kaons to $0.4 < p_t < 0.8$ GeV/c and of protons to $0.6 < p_t < 1.1$ GeV/c. In addition, only particles in the rapidity window $|Y|<0.5$ are selected.
Contamination from misidentified and secondary particles is corrected for since it reduces the correlation strength which is to first order equivalent to increasing
the source size. 

\begin{figure}[ht]
\vspace{-25pt}
\par\centering
\resizebox*{.85\textwidth}{!}{\includegraphics*{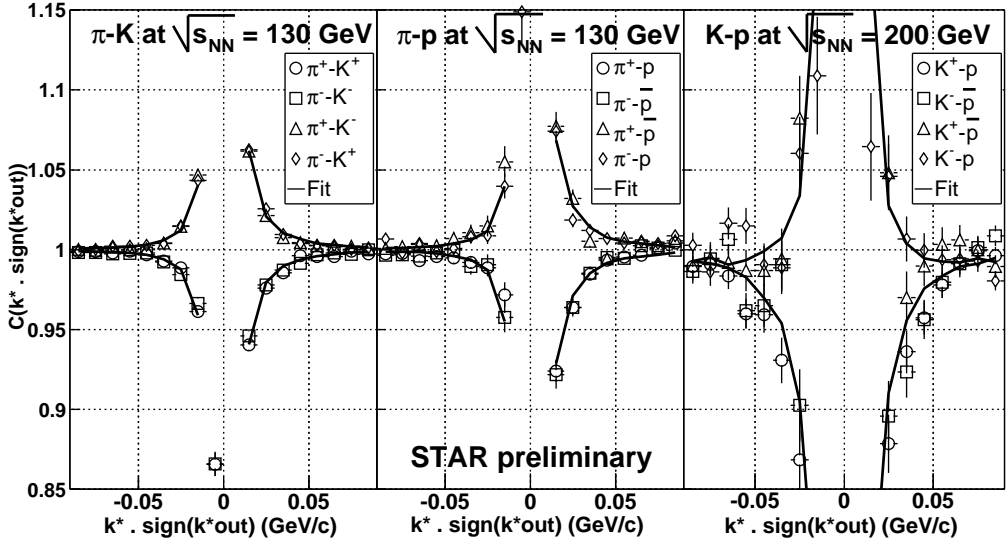}}

\vspace{-30pt}
\caption[Fig1]{\centering \label{Fig1} 
$\pi-K$, $\pi-p$ and $K-p$ correlation functions. 
 }
\vspace{-25pt}
\end{figure}

$\pi-K$, $\pi-p$ and $K-p$ correlation functions are shown in figure 1.  
A clear asymmetry between $k^*_{out}>0$ and $k^*_{out}<0$ can be seen
in the $\pi-K$ and $\pi-p$ correlation functions while the $K-p$  correlation
functions are symmetric within statistical errors. This indicates that pions
are emitted at a different average space-time point than kaons and protons. To
extract quantitative estimates of the source size and asymmetry, a fit
is performed. Correlation functions are calculated following the
method described in Ref ~\cite{LedLubo}.  The calculation is performed on reconstructed pairs in order to match the experimental acceptance. Positions
are assigned to particles forming these pairs from a three dimensional Gaussian distribution,
representing the particle separation in the pair rest frame.
The free parameters are $\langle \Delta r^*_{out} \rangle$ 
($\langle r^*_{out}(\pi)-  r^*_{out}(K) \rangle$ , $\langle r^*_{out}(\pi)-  r^*_{out}(p) \rangle$, or $\langle r^*_{out}(K)-  r^*_{out}(p) \rangle$ ) 
, the mean of the
Gaussian in the direction parallel to the pair transverse momentum and
$\sigma$ the width of the Gaussian that is set to the same value in all
3 dimensions. 
The parameters that best fit simultaneously the four correlation functions
constructed for each kind of pairs are summarized in table 1.
$\langle \beta _t \rangle$ is the average pair transverse velocity at low $k^*$.
The corresponding
calculated correlation functions are shown in figure 1.

\begin{table}
\centering
\begin{tabular}{ccccc}
\hline
\hline
\multicolumn{1}{c}{pair}
&\multicolumn{1}{c}{$\sigma$ (fm)}
&\multicolumn{1}{c}{$\langle \Delta r^*_{out} \rangle$ (fm)}
&\multicolumn{1}{c}{$\chi^{2}$ / dof}
&\multicolumn{1}{c}{$\langle \beta_{t} \rangle$}\\
\hline
$\pi-K$ ($\sqrt{s_{NN}} = 130$ GeV) & $12.7 \pm 0.2  _{-3.2}^{+1}$ & $-6.2 \pm 0.4  _{-0.3}^{+1.2} $ & 107.3/64 & 0.72 \\
$\pi-p$  ($\sqrt{s_{NN}} = 130$ GeV)  & $12.0 \pm 0.4 _{-5}^{+2}$     & $-7.1 \pm 1.2 _{1}^{+3}$          & 123/64     & 0.66 \\
$K-p$ ($\sqrt{s_{NN}} = 200$ GeV)   & $8.5 \pm 0.4_{-5}^{+2}$          & $0.9 \pm 0.7 _{-3}^{+3}$          & 107/64     & 0.62 \\
\hline\hline
\end{tabular}
\vspace{5pt}
\\
Table 1.


\vspace{-20pt}
\end{table}

\section{Non-identical particle correlations and transverse flow}

Figure 2 shows a comparison of the data with predictions from the blast wave  
parameterization and the hadronic cascade model, RQMD. In the blast 
wave parameterization, flow introduces the ordering 
$\langle r^*_{out}(\pi)-  r^*_{out}(K) \rangle <0$, $\langle r^*_{out}(\pi)-  r^*_{out}(p) \rangle < 0$ and $\langle r^*_{out}(K)-  r^*_{out}(p) \rangle <0$, which agrees well
with the data within error. The magnitudes of  $\langle r^*_{out}(\pi)-  r^*_{out}(K) \rangle$ and $\langle r^*_{out}(\pi)-  r^*_{out}(p) \rangle$ are also well reproduced, which is remarkable since no parameters
were tuned to reproduce this data. It is especially interesting to notice that no
additional differences between the emission time of pions, protons and kaons
are needed to explain the data. 
RQMD reproduces the three different shifts between pion, kaon and
proton sources. It has also been found to agree with the asymmetry between pion and protons sources measured in Pb-Pb collisions at $\sqrt{s_{NN}}=$ 17.3 GeV
~\cite{NA49}.  In RQMD, flow builds up through hadronic rescattering, which
yield to the same ordering found with the blast wave parameterization.
In addition,  resonance feed-down and differences in hadronic cross sections 
between pions, kaons and protons, introduce a time ordering in the emission
of pions, kaons and protons. On average, kaons are emitted first, then pions, and then protons.
Since $\Delta r^*_{out} =  ( \Delta r_{out} -  \beta_t c \Delta t) / \sqrt{1-\beta_t^2}$,
spatial shift and time shift add up in the $\pi-K$ case while they compete against
each other in the $\pi-p$ and $K-p$ cases. However, for all three pair types, the spatial shift due to transverse flow is larger than the time shift.
For example, protons are emitted on average latter than pions due to the very large feed down from $\Delta$ resonances, but this delay is counterbalanced by a larger spatial shift from flow, which is driven by the very large pion-proton cross section, i.e. the $\Delta$ resonances.  

\begin{figure}[ht]
\vspace{-25pt}
\centering
\resizebox*{.75\textwidth}{!}{\includegraphics*{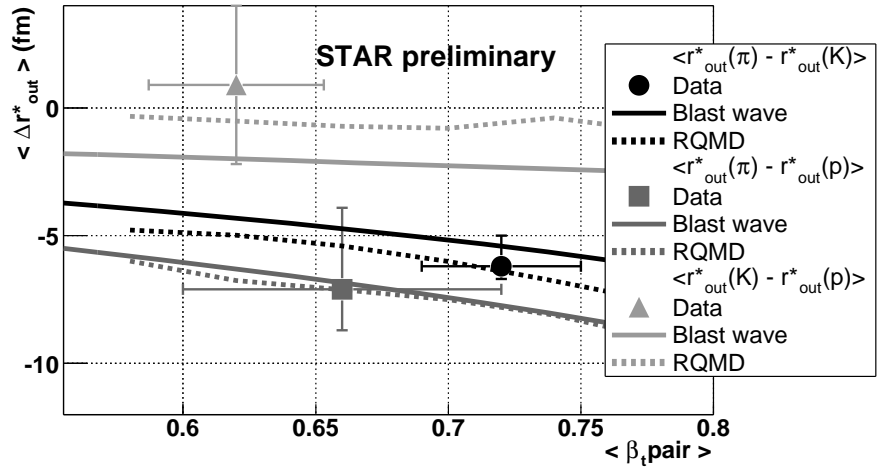}}
\vspace{-35pt}
\caption[Fig2]{\label{Fig2} Average shift between
pion, kaon and proton sources in the pair rest frame. 
\centering}
\vspace{-25pt}
\end{figure}

$\pi-K$, $\pi-p$ and $K-p$ correlation functions have been constructed by
the STAR experiment. They show that pions, kaons and protons are not emitted at the same average space-time position. 
The shifts between the pion, kaon and proton sources are consistent with 
the spatial shift induced by transverse flow as parameterized in the blast
wave framework. 
RQMD also reproduces the shifts primarily due to transverse flow.
Thus, these new results support the picture of a system in collective transverse expansion.

\end{document}